\documentclass[a4paper,conference]{IEEEtran}
\usepackage{todonotes}
\usepackage[utf8]{inputenc}
\usepackage[T1]{fontenc}
\usepackage{microtype}
\usepackage{csquotes}
\usepackage{booktabs}
\usepackage{amssymb}
\usepackage{enumitem}
\usepackage{tikz} 
\usepackage{pgfplots}
\pgfplotsset{compat=1.14}

\usepackage{array}
\newcolumntype{P}[1]{>{\raggedright\arraybackslash}p{#1}}
\newcolumntype{L}[1]{>{\raggedleft\arraybackslash}p{#1}}

\newenvironment{widequotation}{\list{}{\listparindent 1.5em \itemindent\listparindent
		\rightmargin 0pt \parsep 0pt plus 1pt}\item\relax}
{\endlist}

\usepackage{xspace}
\def\signed#1{{\leavevmode\unskip\nobreak\hfil\penalty50\hskip2em
		\hbox{}\nobreak\hfil\raise-3pt\hbox{#1}
		\parfillskip=0pt \finalhyphendemerits=0 \endgraf}}
\newsavebox\mybox
\newenvironment{aquote}[1]
{\savebox\mybox{(#1)}\begin{widequotation}\itshape``\ignorespaces}
	{\unskip"\signed{\usebox\mybox}\end{widequotation}}
	
\newcommand{\inlinequote}[1]{``\emph{#1}''}

\begin{document}

\title{Cross-project Classification of Security-related Requirements}

\newif\ifanonymous
\ifanonymous
\else
\author{\IEEEauthorblockN{Mazen Mohamad, Jan-Philipp Steghöfer, and Riccardo Scandariato}
\IEEEauthorblockA{Chalmers $|$ University of Gothenburg, Gothenburg, Sweden\\ firstname.lastname@gu.se}}
\fi

\maketitle

\begin{abstract}

We investigate the feasibility of using a classifier for security-related requirements trained on requirement specifications available online. This is helpful in case different requirement types are not differentiated in a large existing requirement specification. Our work is motivated by the need to identify security requirements for the creation of security assurance cases that become a necessity for many organisations with new and upcoming standards like GDPR and HiPAA.
We base our investigation on ten requirement specifications, randomly selected from a Google Search and partially pre-labelled.
To validate the model, we run a 10-fold cross-validation on the data where each specification constitutes a group.

Our results indicate the feasibility of training a model from a heterogeneous data set including specifications from multiple domains and in different styles. However, performance benefits from revising the pre-labelled data for consistency.
Additionally, we show that classifiers trained only on a specific specification type fare worse and that the way requirements are written has no impact on classifier accuracy.

\end{abstract}



\section{Introduction}
\label{sec:introduction}

Assurance Cases are documented bodies of evidence used to reason about certain properties of a system~\cite{sac}. Such cases have been widely used in safety-critical domains such as automotive, aviation, and medical devices as a framework for safety assurance. Using assurance cases for security is a relatively new development which is quickly gaining a foothold.

One reason for this is that cyber-security has been gaining attention in the past few years in safety-critical domains. In the automotive domain for instance, manufacturers started to increase their security efforts when two hackers succeeded to remotely hack a Jeep Cherokee and were able to control essential parts of the vehicle such as the steering wheel~\cite{jeepHack}. Such unauthorised access compromises the safety of the vehicle's passengers as well as that of other vehicles on the road. 
Authorities are also acting to make cyber-security an area of focus when it comes to safety-critical systems. The new ISO standard 21434~\cite{iso21434}, e.g., states that car manufacturers must provide structured argumentation and evidence for their claims regarding the security of their systems. This will happen in the form of security assurance cases (SAC), following an approach similar to the one long practiced in the area of safety.

SAC will not only be created for new projects, but will also need to be prepared retrospectively for existing systems. As an assurance case always starts with a set of relevant requirements, it will be necessary to identify all security-related requirements for such systems. A typical safety-critical system such as a vehicle, an aircraft, or an insulin pump can be based on requirement specifications with thousands of requirements spread over several documents~\cite{cnn,firesmith2003}. In many cases, security-related requirements are not marked as such and thus a manual search is necessary, incurring a potentially very high cost for identifying them manually.

The alternative is that companies mine large amounts of data in different repositories to find security-related requirements that can be used as claims in SACs in an automated way. Machine learning approaches and in particular classifiers can support such a task, but require a robust training set in order to be able to classify security-related requirements with high accuracy. Constructing such a training set manually can again incur high cost.

In this paper, we investigate the usefulness of training a classifier on existing, labelled projects freely available on the internet. If such a classifier provides sufficient accuracy, companies will be able to save significant effort and time. These cost reductions depend on the quality of the classification, the manual effort required to find suitable projects for learning, and how closely the classification results need to be checked and revised. Our first research question is thus:

\textbf{RQ1:} Which performance can a classifier for security-related requirements achieve if it is trained on data from other projects?

While it is desirable that practitioners can use training data without much additional effort, previous work shows that classification of security requirements~\cite{knauss2011} or non-functional requirements in general~\cite{kurtanovic2017} is difficult across projects. We thus also explored the impact of the structure of the requirements and the specification types. Our second research question is therefore:

\textbf{RQ2:} Which impact do factors such as the structure of the requirements and the type of requirement specification have on classifier performance?

We answer these research questions by performing a 10-fold cross-validation with ten different requirements specifications mined from the Internet where each of the specifications forms one group. We also manually improve the labelling of the requirement specifications and explore different modifications to this general setup.

Our results show that performance of a classifier trained on nine of the specifications and applied to the remaining one can be very good, but that caveats apply with regard to the data quality of the training data. We thus also consider performance of the classifier when the training set has been revised manually. We find that the structure of the requirements and the type of the specification play little role and that, on the contrary, heterogeneous training data is preferable. Based on these findings, we make suggestions for how to apply the approach in practice and discuss its limitations.

\section{Background and Related Work}
\label{sec:related-work}


\subsection{Security-related Requirements}
Cyber-security (referred to simply as security in the remainder of this article) in general is defined as:

\begin{aquote}{Scatz et al.~\cite{schatz2017}}
The approach and actions associated with security risk management processes followed by organizations and states to protect confidentiality, integrity and availability of data and assets used in cyber space. The concept includes guidelines, policies and collections of safeguards, technologies, tools and training to provide the best protection for the state of the cyber environment and its users.
\end{aquote}

The triad \textbf{c}onfidentiality, \textbf{i}ntegrity and \textbf{a}vailability (CIA) is considered to represent the important security properties~\cite{samonas2014}. 
The requirements that directly target one of these properties are security requirements. Additionally, the requirements to comply with a certain security regulation, standard, or best practice, e.g., ISO 27001~\cite{iso27001} are also counted as security requirements.
One challenge of requirements engineering for security is that requirement engineers often lack expertise in security~\cite{firesmith2003}. Hence they might not be able to correctly tag security requirements in case they are not explicit.
The motivation for this study is to support constructing security arguments for security assurance cases. Hence we are interested in identifying not only direct security requirements, but also ones that relate to CIA, such as the creation of a log to keep track of a user's actions. We consider these to be \emph{security-related requirements}. All security requirements are thus also security-related.

\subsection{Security Assurance Cases}
\begin{figure}[!t]
    \centering
    \includegraphics[width=1\linewidth]{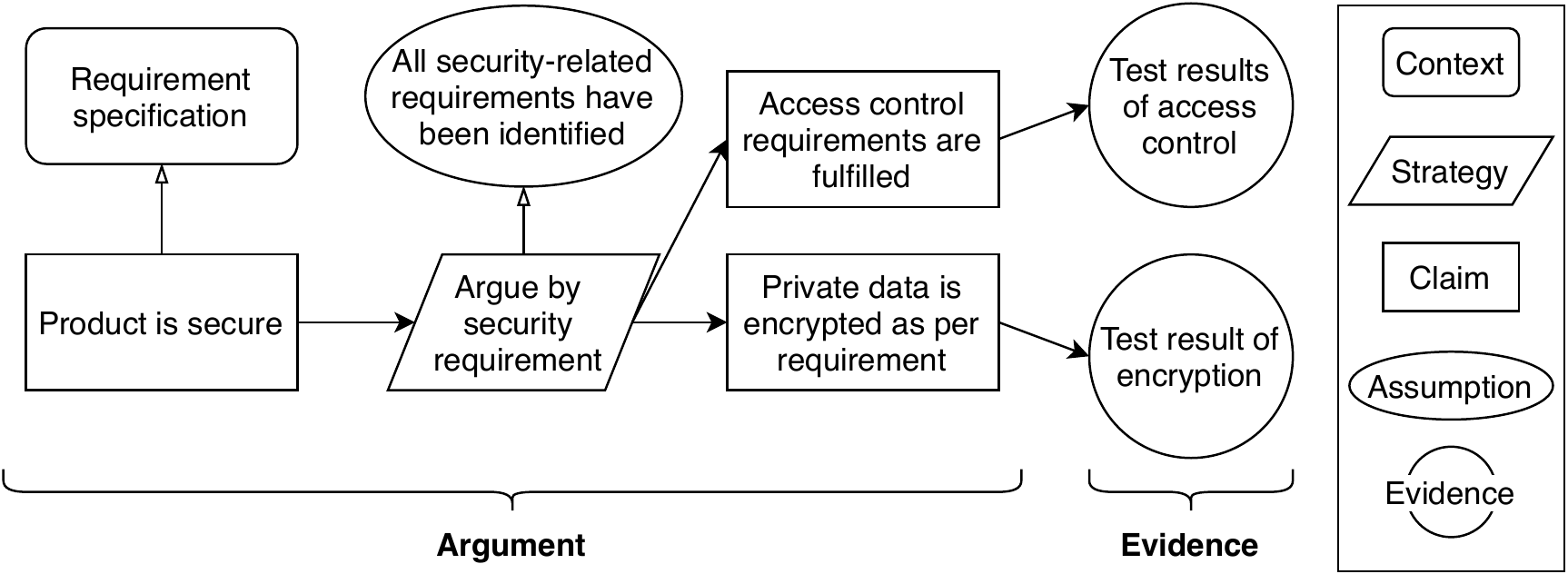}
    \caption{Structure of a Security Assurance Case}
    \label{fig:SAC}
\end{figure}

Assurance Cases (AC) are defined as  \inlinequote{A reasoned and compelling argument, supported by a body of evidence, that a system, service or organisation will operate as intended for a defined application in a defined environment.} by GSN~\cite{GSN_standard}. A \textbf{s}ecurity \textbf{a}ssurance \textbf{c}ase (SAC) is a special type of an AC where security is the quality property of focus~\cite{alexander2011,goodenough2007}. An SAC consists of two main parts: the argument and the evidence, as shown in Figure~\ref{fig:SAC}. The figure also shows the main elements of an SAC: the \emph{(i) claim}, which represents a security goal of the artefact for which the SAC is built; the \emph{(ii) context}, in which the claims apply; the \emph{(iii) strategy}, which is the driver for breaking down a claim into sub-claims; the \emph{(iv) assumptions}, in which the assumptions made while creating the argument are made explicit; and the \emph{(v) evidence}, which is an artifact that justifies a certain claim. 

There are multiple options for approaching the creation of an SAC argument. One common approach is to use security requirements as the starting point~\cite{38_agudo2009,54_haley2005,55_calinescu2017}. An example is shown in Figure~\ref{fig:SAC}. To ensure the completeness of the case, all relevant security-related requirements have to be addressed. Not meeting that criterion negatively impacts the quality of the case. Hence, it is important to identify explicit security requirements as well as security-related requirements in a given requirements specifications document.

\subsection{Automatic Classification Tasks}
Automatic classification tasks address the problem of arranging data into groups or classes. There are two main types of classifiers problems: binary (when we only have two classes), and multi-class (when we have more than two classes). Classifiers learn from data sets which are usually called training sets. A classifier extracts features from the training sets that it uses two differentiate the classes. There are two main types of learning: supervised, and un-supervised. Supervised learning, which is used in this study, uses a labelled training set that contains examples for each class. The classifier uses a Machine Learning (ML) algorithm to learn patterns based on the features from the training set which allows it to predict the label of unlabelled data (often referred to as test data)~\cite{friedman2001}. 

\subsection{Related Work}
\subsubsection{Support for SAC construction}
Machine learning has not been used for constructing security assurance cases before. However, some studies have focused on how to extract relevant information for constructing SACs from different sources such as document repositories and models.

Chindamaikul et al.~\cite{docRet} conducted a study which resulted in an approach to extract information from a large data set using document retrieval and formal concept analysis techniques. In contrast our study in which users get relevance-predictions for a set of requirements, Chindamaikul et al.~\cite{docRet} suggest that the user writes a query with the desired keywords and gets the relevant documents back. 
Hawkins et al.~\cite{hawkins2015} explore the possibility for a model-based approach to construct SAC based on information extracted from models. 
Shrott and Weber~\cite{shrott2015} created a tool for dynamically analyzing code coverage for SAC evidence creation. 
Tippenhauer et al.~\cite{tippenhauer2014} suggest an approach for automatically creating security argument graphs. The approach takes different documents as input, such as security goals, attacker model, and system description. However, there is no discussion on how to extract security-related data from a more generic document such as a system description.

\subsubsection{Classification of requirements}
Automatic classification and categorization of requirements has been explored in multiple studies. 
Researchers have suggested approaches for classifying functional requirements in order to analyse customers' requirements~\cite{xu2007} and to classify requirements documents into content topics in order to assist reviewers from certification authorities in finding inconsistencies~\cite{ott2014}.
Classifying security requirements has also been reported in literature. It was however targeting other goals than supporting the creation of SAC. Knauss et al.~\cite{knauss2011} approach the problem that security issues are ignored early in software projects by creating an early indication of security issues based on security requirements. The approach uses a classifier to classify security related requirements. The researchers evaluate the approach by using three requirement specifications from industry with a total of 510 requirements, of which 187 are security related. The paper also tests transferability of a model trained on two specifications to the third. The results showed an F${}_1$\-measure below $50\%$ in these cases, while a cross validation on the whole data set gave an average ${}_1$\-measure of $84\%$.

Kutranovic et al.~\cite{kurtanovic2017} also study the classification of requirements, but into different classes, namely functional, and non-functional. The researchers also study a multi-class classification of the non-functional requirements. One of these classes is security. The data which Kutranovic et al.\ used was from the RE17 conference data challenge. The results of security requirements classification show an F${}_1$\_measure of $74\%$. A similar study to classify non-functional requirements is done by Cleland-Huang et at.~\cite{cleland2007}. The data was collected from 15 projects with a total of 684 requirements, of which 326 are non-functional. The results show an average recall of around $70\%$. However, the score for security requirements was below that average.

In our study, we use a heterogeneous dataset with significantly more requirements than the related work, as our aim is study the possibility of cross-project training and prediction. We also discuss patterns in which the classifier fails to predict correctly.

\section{Method}
\label{sec:method}

 \begin{figure*}[tb]
    \centering
    \includegraphics[width=1\linewidth]{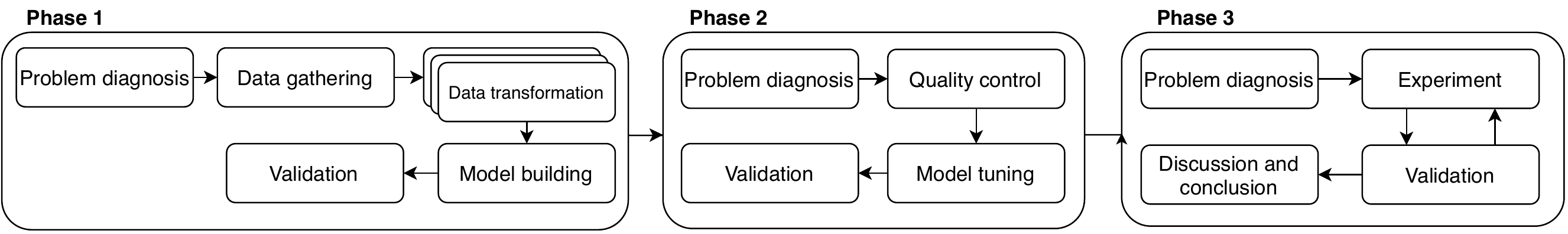}
    \caption{The three-phased methodology we followed to design and validate the cross-project classification of security requirements.}
    \label{fig:method}
\end{figure*}

We followed a three-phased iterative approach as shown in Figure \ref{fig:method} to answer our research questions. We chose this problem-solving paradigm since it allows us to create and apply artefacts in an iterative manner with design and validation delivering insights that can be used for improvement in the next iteration.

\subsection{Phase 1}
In this phase, we focused on building the data set for this study, preparing it for training the classifier, and building and training a first version of the classifier. We describe the steps we performed in this phase in the following subsections.

\subsubsection{Data Gathering}
We identified ten different requirement specifications from commercial projects, student projects, and research projects with a total of 3003 requirements for use in this study. 
The data set was collected by performing a Google search. Our goal was to obtain data that was as  heterogeneous as possible. Hence, we did not want to restrict the search on a specific domain or a specific type of requirement specifications. The only restriction we had was on the file type, as we were only interested in file types from which we can easily extract the data.  The search string we used was:
\\

\texttt{requirements specifications filetype:xls OR xlsx OR pdf}\\

We went through the results and excluded files from which extraction of data was not possible, e.g., due to access restrictions or non-English text. We then selected the first ten files that fit our criteria for inclusion in our data set.

\begin{table*}[tb]
\centering
\caption{Publicly available requirements specifications included in the study.}
\begin{tabular}{@{}llllllll@{}}
 \toprule
 \multicolumn{3}{l}{} & \textbf{Specification} & \textbf{Total number of} & \multicolumn{2}{l}{\textbf{Security requirements}} & \\
 \textbf{Spec} & \textbf{Project} &\textbf{Project type} & \textbf{type} & \textbf{requirements} & \textbf{Phase 1} & \textbf{Phase 2} & \textbf{Pre-labeled}\\
 \midrule
 1 & Mobile application for restaurants & Student project & SRS &   98& 20 & 18 & Yes\\
 2 & University inventory management & Student project & SRS &   27 & 9  &11 & Yes\\
 3 & Financial management system    & Commercial project & RFP &180 & 110 &  103 & Partially \\
 4 & Research data management system & Commercial project &SRS & 92 & 8 & 9 & Partially \\
 5 & Software development platform & Research project & BL & 567& 68 & 104 & Partially \\
 6 & Financial management system & Commercial project & RFP &171 & 41 & 41 & No\\
 7 & Financial management system & Commercial project & RFP &995 & 5 & 179 & Partially \\
 8 & Customer relation management system & Commercial project & RFP &127& 25 &  38 & Yes\\
 9 & Electronic document management system   & Commercial project &BL & 253 & 22 & 34 & No\\
 10 & Human resources management system  & Commercial project & RFP & 493 & 70 & 57 & Yes\\
 \midrule
 \multicolumn{4}{@{}l}{\textbf{Total}}  & \textbf{3,003} & \textbf{378} & \textbf{594} &   \\
 \bottomrule
\end{tabular}
\label{tab:method:projects}
\end{table*}

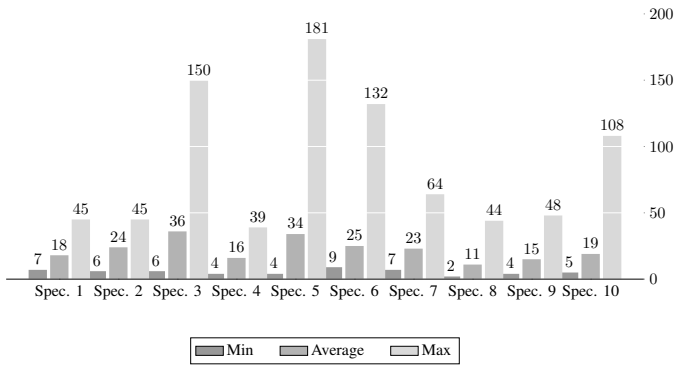
\begin{figure}
	\centering
\resizebox{\linewidth}{!}{%
\begin{tikzpicture}

  \begin{axis}[
        ybar, axis on top,
        height=8cm, width=15.5cm,
        bar width=0.4cm,
        ymajorgrids, tick align=inside,
        major grid style={draw=white},
        enlarge y limits={value=.1,upper},
        ymin=0, ymax=200,
        axis x line*=bottom,
        axis y line*=right,
        y axis line style={opacity=0},
        tickwidth=0pt,
        enlarge x limits=true,
        area legend,
        legend style={
            at={(0.5,-0.2)},
            anchor=north,
            legend columns=-1,
            /tikz/every even column/.append style={column sep=0.5cm}
        },
        symbolic x coords={
           Spec.~1,Spec.~2,Spec.~3,Spec.~4,
           Spec.~5,Spec.~6,Spec.~7,Spec.~8,
           Spec.~9,Spec.~10},
       xtick=data,
       nodes near coords={
        \pgfmathprintnumber[precision=0]{\pgfplotspointmeta}
       }
    ]
    \addplot [draw=none, fill=gray!80] coordinates {
      (Spec.~1,7)
      (Spec.~2,6) 
      (Spec.~3,6)
      (Spec.~4,4) 
      (Spec.~5,4) 
      (Spec.~6,9)
      (Spec.~7,7) 
      (Spec.~8,2)
      (Spec.~9,4)
      (Spec.~10,5) };
   \addplot [draw=none,fill=black!30] coordinates {
      (Spec.~1,18)
      (Spec.~2, 24) 
      (Spec.~3,36)
      (Spec.~4,16) 
      (Spec.~5,34) 
      (Spec.~6,25)
      (Spec.~7,23) 
      (Spec.~8,11)
      (Spec.~9,15)
      (Spec.~10,19) };
   \addplot [draw=none, fill=gray!30] coordinates {
      (Spec.~1,45)
      (Spec.~2, 45) 
      (Spec.~3,150)
      (Spec.~4,39) 
      (Spec.~5,181) 
      (Spec.~6,132)
      (Spec.~7,64) 
      (Spec.~8,44)
      (Spec.~9,48)
      (Spec.~10,108) };

    \legend{Min,Average,Max}
  \end{axis}
  \end{tikzpicture}}
\caption{Number of words in the requirements per specification} \label{fig:numReq}
\end{figure}

Table~\ref{tab:method:projects} shows information about the requirement specifications from which we collected our data. As the table indicates, the documents are from different projects and contain different numbers of requirements. Some of the documents stem from commercial projects, one has been created by researchers, and yet others have been created by students in university projects.
There is also a difference in the specification type, as in some cases they are System Requirement Specifications (SRS), and in other cases they are Requests for Proposals (RFP), or items of the Backlogs (BL) of certain products. The requirements in different documents are written on different levels, e.g., domain level, user level, and system level. Hence, there are different levels of details in the requirements, which makes the length (number of words) vary significantly among them. Figure~\ref{fig:numReq} shows the minimum, maximum, and average number of words per requirement for each of the 10 projects. As can be seen from the figure, all projects have very short requirements (less than 10 words). However when it comes to the average, the range is between 11 and 34 words. The difference is even bigger when we look at the longest requirements which vary between 39 and 181 words. Additionally, the different documents contain requirements written in different styles. 
Some are written as user stories, some as instructions to potential vendors, and others in a conversational language style.

\subsubsection{Labelling}
\label{sec:method:labelling}
In Phase 1, we manually assigned labels to the requirement specifications that contained no labelling or only a partial labelling as we were planning to use a supervised learning algorithm which requires labelled data. 
We differentiate two classes for the labelling: \emph{(i)} security requirements are the \emph{positive class}; \emph{(ii)} all other requirements are the \emph{negative class}.
We consider all requirements that address security properties of a system to be security requirements. This includes but is not limited to requirements for access control, data confidentiality, system availability, data integrity, and logging.

Many of the requirements documents used in this study had labelled requirements. How requirements were labelled differed between the specifications. In some cases, different types of requirements were in separate sheets, in others tags were used. We unified the representation when preparing our data files. However, half of the labelled documents did not have specific labels for security, but rather a label to indicate non-functional or quality requirements (marked as \emph{Partially} labelled in Table~\ref{tab:method:projects}). Hence, we performed manual labelling for such specifications. The complexity of this task varied depending on the project and how the requirements are written. This is due to the fact that some requirements were written as functional requirements but contained \emph{implicit} security requirements. An example of such a requirement is: 

\begin{aquote}{Specification 7}
The system shall provide the ability for an authorized user to reissue a payment according to configurable business rules e.g., supervisor approval
\end{aquote}

This requirement is functional but also restricts access to a function to authorized users, which makes it a security-related requirement. To determine to which class such requirements belong, we followed these steps:
\begin{itemize}
    \item Extract the security-related part of the requirement (in our example case that would be restricting access to the functionality of reissuing a payment to authorized users).
    \item Search the document for a different requirement which covers the extracted security-related part.
    \begin{description}
	    \item[If found,] we consider the original requirement to be not security-related. In this study, this was rarely the case.
		\item[Else,] we consider the original requirement to be security-related and label it correspondingly.    
    \end{description}

\end{itemize}

\subsubsection{Data Pre-Processing}
\label{sec:results:cleanup}
To prepare data for the classifier, we pre-processed the requirements specifications following standard procedures for machine learning. We used the machine learning library scikit-learn~\cite{scikit-learn} for  the entire study.

\begin{LaTeXdescription}
    \item[Remove noise:] We removed punctuation and special characters by defining regular expressions using the \emph{re} library in python, and applying them on the data. 
    \item[Extract words:] Since the requirements are written in a natural language, the words are separated with spaces. Hence, we considered every sequence of characters between two spaces to be a word.
    \item[Ignore case:] Since case sensitivity is not important in our problem, we converted all words to lower case.
    \item[Remove stop words:] Stop words refer to the most commonly used terms in a language. These words usually appear in most sentences and thus do not contribute to classifying the requirement. We used the pre-defined \emph{English} stop words list in scikit-learn. 
    \item[Stemming:] To ensure that words that stem to the same root, e.g., ``authorization'' and ``authorize'', are treated the same way, we used the library PorterStemmer in scikit-learn to perform the stemming task.
\end{LaTeXdescription}

\subsubsection{Feature extraction}
In this study we use the Bag of Words (BOW) representation of features, where each word of the textual corpus is considered a feature. In particular, we used TF-IDF (Term Frequency-Inverted Document Frequency) which is a statistical measure that reflects how important a specific term is in a given corpus~\cite{manning2008introduction}. Term frequency gives weight to a term that appears frequently in a given document (requirement in our case):
    \begin{equation}
         TF(t)=\frac{\mbox{\emph{Number of times term t appears in a requirement}}}{\mbox{\emph{Total numbers of terms in the requirement}}}
    \end{equation}
    
Inverted document frequency gives weight to terms that are rare in the documents.

    \begin{equation}
         IDF(t)=\log_{e}(\frac{\mbox{\emph{Total number of requirements}}}{\mbox{\emph{Number of requirements that include t}}}) 
    \end{equation}

In our case, we expect the terms relevant to the positive class to be rare among all the requirements as the positive class is significantly smaller than the negative one. Hence, we expect TF-IDF to have good performance.

    The final score is the multiplication of the two values:
    \begin{equation}
         TF-IDF_{score} = TF * IDF
    \end{equation}


\subsubsection{Up-sampling}
Since we had significantly more non-security requirements than security ones in the data set, we used an up-sampling technique to achieve balanced classes in the training set for the machine learning model. After splitting the data into training and test sets, we applied the Synthetic Minority Oversampling (SMOTE) technique~\cite{chawla2002smote} on our training set. 
Rather than replicating the minority observations, SMOTE creates synthetic records based on the existing minority records.

\subsubsection{Selection of Classification Algorithm}
Our problem is a binary classification to predict whether a textual requirement belongs to the positive class (security-related requirement), or the negative class (non-security related requirement). The selection of the algorithm was based on an initial test to compare the algorithms on our data set. We tested three of the most commonly used algorithms for text classification: Random Forest (RT) \cite{ho1995random}, Linear Support Vector Machine (Linear SVM) \cite{platt1998svm}, and Naive Bayes (NB) \cite{maron1961nb}. The results of this test showed a slight advantage of RT, so we decided to use it for this study.

\subsection{Phase 2}
In this phase, we focused on enhancing the results from the previous phase and extracting information that could help us to build better classifiers in the future. We started by diagnosing the reasons for the underwhelming results of Phase 1, especially in the cross-project validation step. 

\subsubsection{Quality assurance of labeling}
A significant issue we identified were incorrectly pre-labelled requirements.
We thus decided to revise the labelling of all 3003 requirements in the data set. The most common modification of the original labelling was to move requirements from the negative class to the positive class by applying the decision procedure described in Section~\ref{sec:method:labelling}. As a result, the number of security related requirements increased from 378 to 594.  Table~\ref{tab:method:projects} shows the changes in the number of security requirements from Phase 1 to Phase 2.

For some specifications, the number of security requirements in a certain specification decreased during the quality check. This happened when security-related requirements (e.g., to create audit logs in specification 10) were followed by refined requirements (e.g., on what the logs should look like and how they should be sorted). The pre-labelled data marked all of these as security-related, while our labelling removed the label from the follow-up requirements.

\subsubsection{Improving model hyper-parameters}
The results of the cross-project validation suggest that we might have an over-fitting issue. To eliminate that risk, we tuned the following hyper-parameters of the random forest algorithm: 
\begin{LaTeXdescription}
	\item[(i) n\_estimator] indicates the number of trees in the model (the more trees, the lower the risk of over-fitting);
	\item[(ii) max\_features] defines how many features each tree in the ensemble is randomly assigned; and
	\item[(iii) max\_depth] specifies the maximum depth of each tree.
	\item[(iv) bootstrap] indicates whether the whole dataset will be used for building trees, or only bootstrap samples of the data.
\end{LaTeXdescription}

We used scikit-learn's \texttt{RandomSearchCV} function which performs a random search of the hyper-parameters and performs k-fold cross-validation for each set of parameters. The random search was performed for 100 iterations and results were evaluated with a three-fold cross-validation.

\subsubsection{Feature extraction}
In this step, we were interested in extracting the most important features that the classifier used. This information gave insights into how the classifier interpreted the different specifications. We used the \texttt{feature\_importances\_} attribute returned by the classifier after the training. This list returns the features ordered by importance, which is represented by a value. The sum of the importance of all features sum to 1. When represented in this study, we multiply the importance values by 100 to present a percentage of the overall importance.

\subsection{Phase 3}
In order to better understand the results from phases 1 and 2, we added a third phase in which we ran specific experiments on the data, in particular to investigate the impact of certain optimisations. In particular, we were interested in the following questions:

\begin{itemize}
	\item Which impact does the optimisation of the hyper-parameters have on the results?
	\item Can an extended list of stop words improve the results?
	\item Which impact do the different specification types have?
\end{itemize}

To answer these questions, we retrained and reclassified our data with different combinations of specifications and with different parameters.

To answer the first question, we treated the hyper-parameters as an independent variable, and created two treatments, one using the default values of the hyper-parameters (used in Phase 1) and the second using the values of the parameters which we got from the random search cross-validation (used in Phase 2). We again ran the experiment as a 10-fold cross-project validation and  used the labelling from Phase 2. The dependent variables used were the accuracy, precision, and recall of the classification. We report the results as the difference between the treatments for each of our dependent variables, in each fold, and calculate the averages.

To answer the second question, we added terms which often appear in requirements, such as \emph{user}, \emph{shall}, and \emph{vendor} to the stop-word list and checked if this would improve the classification results. We used the stop-word list as our only independent variable in this experiment. We created two treatments, one using the default English language stop-words list used in phases 1 and 2 and another using an extension to that list by the three mentioned terms. We ran the experiment as a 10-fold cross-validation. We used the labelling from Phase 2. The dependent variables used were the accuracy, precision, and recall of the classification. We report the results as the difference between treatment one and two for each of our dependent variables in each fold and calculate the averages.

To investigate the last question, we hypothesised that the structure of the requirements does not make a significant difference to the classifier. To test that, we conducted an experiment with one independent variable, which is the dataset used for classification. We created two treatments, one using the dataset from Phase 2, and another using the subset that contains only requirements from specifications with specification type RFP (five specifications). We ran the experiment as a 5-fold cross-validation. The dependent variables used were the accuracy, precision, and recall of the classification. We report the results as the values of the dependent variables for each project, and discuss the difference to the results of Phase 2. 

\subsection{Threats to validity}

In terms of \textbf{internal validity}, we consider the data labelling and the selection of the algorithm.

We \emph{labelled the data set} at two different points in time: once before Phase 1 for the unlabelled specifications; and once in Phase 2 for the whole data set. The labelling was done by one person, hence there is the risk that judgement was subjective. To mitigate this risk, an additional person was asked to label a sample of the data. An agreement analysis using Cohen's Kappa~\cite{kappa} showed an agreement of over $80\%$ which is considered a good agreement value~\cite{kappa1,kappa2}.

 The \emph{selection of the RF algorithm} was based on a preliminary run which compared three algorithms. There is a possibility that another algorithm had performed better if the data cleaning and hyper-parameters tuning had been performed before comparing. We also did not test other feature extraction models than BOW. However, the purpose of this study was not to find an optimal approach, but rather to investigate if cross-project validation was feasible.
 
In terms of \textbf{external validity}, we consider \emph{overfitting} and \emph{imbalanced data sets}.

 \emph{Overfitting} is a common problem in machine learning~\cite{hawkins2004}. It causes the classifier to perform very good during training and optimization, but very poorly when applied. To avoid this issue, we performed a hyper-parameter tuning with a cross-validation random search with 300 fits in Phase 2.
 
We used oversampling to tackle the issue of \emph{imbalanced data sets}. This might have increased the likelihood of overfitting according to Sotiris et al.~\cite{imbalanced}. To reduce this risk, we used the synthetic minority oversampling technique rather than random oversampling and applied it after splitting the data set into training and testing data.

\section{Results}
\label{sec:results}


\subsection{Validation -- Phase 1}
\label{sec:results:phase1}
We performed an initial validation of the classification approach by using a standard 80/20 split.
This means that we split the data randomly into a training set ($80\%$ of the data) and a testing set ($20\%$ of the data).  Table~\ref{ResTab1:fullDataValidation} shows the results of the test. We obtained an average accuracy of $94.9$. However, since the negative class dominates with almost $76\%$ of the data not related to security, this measure is not a sufficient indicator of quality. 
Hence, we also looked at the precision and recall of the results. While precision is generally good at $93$ (i.e., few requirements are marked as security-related even if they are not), recall is low at $58.8$ (i.e., many security-related requirements are not found).
While these results were not stellar, they showed at least that our dataset was viable.

Additionally, we report three different f-measures: the F${}_1$-score is the harmonic mean of precision and recall; in the F${}_{0.5}$-score, precision is given double the weight of recall; and the F${}_2$-score weighs recall higher than precision. Which of these scores is suitable depends on the situation. If false negatives are to be avoided, a focus on recall and thus the F${}_2$ score is a good choice; if false positives are to be avoided, the focus should be on precision and thus the F${}_{0.5}$ score.

\begin{table}[tb]
\centering
\caption{Results of validation using the full data set and an 80/20 split.}
\label{ResTab1:fullDataValidation}
\begin{tabular}{@{}p{0.7cm}p{1cm}p{1cm}p{1cm}p{0.7cm}p{0.7cm}p{0.7cm}@{}}
\toprule 
\multicolumn{4}{l}{} & \multicolumn{3}{l}{\textbf{f-measures}} \\
\textbf{Phase} & \textbf{Accuracy} & \textbf{Precision} & \textbf{Recall} & \textbf{f${}_1$} & \textbf{f${}_{1/2}$} & \textbf{f${}_2$}\\
\midrule
1 & 94.9 & 93 & 58.8 & 72.1 & 83 & 62.7 \\
2 & 94.5 &     87.6 & 80.2  & 83.7 & 86 &  81.6     \\
\bottomrule
\end{tabular}
\end{table}

The next step was to test the quality of the classifier when performing cross-project classification. For this purpose, we used 10-fold-cross-validation, with each requirement specification being one group. We thus trained the classifier on nine of the specifications and tested it on the tenth. This was run ten times so that all combinations were explored.

The results of this experiment are shown in Table~\ref{tab:results:crossproject-phase1}. While some requirement specifications showed good accuracy and precision, recall was generally low and there were significant outliers (marked in \textbf{bold face}) in the table.

\begin{table}[tb]
\centering
\caption{Cross-project validation results of Phase 1. Outliers are marked in \textbf{bold face}.}
\label{tab:results:crossproject-phase1}
\begin{tabular}{@{}p{0.6cm}p{1cm}p{1cm}p{1cm}p{0.7cm}p{0.7cm}p{0.7cm}@{}}
\toprule
\multicolumn{4}{l}{} & \multicolumn{3}{l}{\textbf{f-measures}} \\
\textbf{Spec.} & \textbf{Accuracy} & \textbf{Precision} & \textbf{Recall} & \textbf{F${}_1$} & \textbf{F${}_{1/2}$} & \textbf{F${}_2$}\\
\midrule
1&91.8 & 91.7 & 61.1 & 73.3 & 83.4 & 65.5 \\ 
2&92.6 & 88.9 & 88.9 & 88.9 & 88.9 & 88.9 \\ 
3&64.4 & 76 & 55.4 & 64 & 70.7 & 58.6 \\ 
4&92.4 & 66.7 & \textbf{44.4} & 53.3 & 60.6 & 47.6 \\ 
5&84.1 & \textbf{45.4} & 54.3 & 49.4 & 46.9 & 52.3 \\ 
6&89.5 & 81.8 & 69.2 & 75 & 78.9 & 71.4 \\ 
7&90.7 & 77.9 & 62 & 72.1 & 74.1 & 64.6 \\ 
8&83.5 & 77.8 & \textbf{45.2} & 57.1 & 68 & 49.3 \\ 
9&94.5 & 75 & 75 & 75 & 75 & 75 \\ 
10&92.1 & 89.7 & 50 & 64.2 & 77.4 & 54.9 \\

\midrule
           $\varnothing$ & 87.56 & 77.09 & 60.55 & 67.23 & 72.39&	62.81   \\
\bottomrule
\end{tabular}
\end{table}

\subsection{Validation -- Phase 2}
\label{sec:results:phase2}

When investigating these outliers, we realized that the labelling of data was inconsistent in the data set. Since we are using multiple requirements specifications documents from multiple sources labelled by different people with different standards, such inconsistencies might be unavoidable. An example is when requirements related to logging user actions were considered as security requirements in some projects, but not in others. This influences the accuracy since the classifier might label a requirement about logging user actions as security-related, but the ground truth shows it is not.

In order to address such issues, we revised the labelling of the requirements specifications. In specification 5, e.g., one of the outliers in Phase 1, the number of security-related requirements increased from $68$ to $104$, an increase of more than $50\%$. This not only changes the outcomes when specification 5 is used to validate the classifier, but also has an influence on other projects when specification 5 is included in the training data.

In addition, we optimised the hyper-parameters of the Random Forest classifier. As described in Section~\ref{sec:method},  a few key parameters determine the performance of the classifier. For a detailed discussion of the impact of hyper-parameter tuning on the performance of the classifier, please see Section~\ref{sec:results:tuning}. The values of the key hyper-parameters as per the results of the cross-validation random search are:
\begin{itemize}
	\item n\_estimators=\emph{400}, which is an increase from the default value \emph{100} used in the first phase;
	\item max\_features=\emph{'sqrt'}, which tells the classifier to use the square root of the number of features when looking for the best split. This value is the same as the default value \emph{auto}; 
	\item max\_depth = \emph{None}, which is the same as the default value;
	\item bootstrap=False, which means that the whole dataset would be used to build each tree. This differs from the default value where bootstrap samples are used for building the trees.
\end{itemize}


The final results of Phase 2 are shown in Table~\ref{tab:results:crossproject-phase2}. Overall, the results have been improved as indicated by the averages in the last row.

\begin{table}[tb]
\centering
\caption{Cross-project validation results of phase 2}
\label{tab:results:crossproject-phase2}
\begin{tabular}{@{}p{0.6cm}p{1cm}p{1cm}p{1cm}p{0.7cm}p{0.7cm}p{0.7cm}@{}}
\toprule
\multicolumn{4}{l}{} & \multicolumn{3}{l}{\textbf{f-measures}} \\
\textbf{Spec.} & \textbf{Accuracy} & \textbf{Precision} & \textbf{Recall} & \textbf{F${}_1$} & \textbf{F${}_{1/2}$} & \textbf{F${}_2$}\\
\midrule
1 & 94.9 & 93.9 & 77.8 & 84.8 & 90.2 & 80.6 \\ 
2 & 92.6 & 90.1 & 90.1 & 90.1 & 90.1 & 90.1 \\ 
3 & 83.3 & 96.1 & 73 & 82.9 & 90.3 & 76.7 \\ 
4 & 96.7 & 85.7 & 77.8 & 82.3 & 84 & 79.2 \\ 
5 & 92.4 & 86.8 & 69.2 & 77 & 82.6 & 72.1 \\ 
6 & 91.2 & 88.2 & 73.2 & 80 & 84.7 & 75.8 \\ 
7 & 94.3 & 82.5 & 86.6 & 84.5 & 83.2 & 85.7 \\ 
8 & 92.9 & 100 & 76.3 & 86.6 & 94.2 & 80.1 \\ 
9 & 95.7 & 92.6 & 73.5 & 82 & 88 & 76.7 \\ 
10 & 95.7 & 91.9 & 65.4 & 76.4 & 85 & 69.4 \\

\midrule
           $\varnothing$  & 89.6 & 90.8 & 76.3 & 82.7 & 87.4 & 78.8 \\ 
\bottomrule
\end{tabular}
\end{table}

\subsection{Validation --- Phase 3}
\label{sec:results:phase3}

We ran a serious of experiments to gain a better understanding of the impact of the hyper-parameters and of different combinations and modifications of our data.

\subsubsection{Hyper-parameters tuning}
\label{sec:results:tuning}

The results of Phase 2 reported above contained two improvements: the manual revision of the data labelling and a tuning of the hyper-parameters of the Random Forest classifier. In order to gain insight in which of these optimisations had the bigger impact, we quantified the changes when running the classifier with standard hyper-parameters and with optimised parameters.

\begin{table}[tb]
\centering
\caption{Changes in quality metrics when tuning the hyper-parameters for each specification.}
\label{tab:results:hyperparameters}
\begin{tabular}{@{}p{0.6cm}p{1.2cm}p{1.2cm}p{1.2cm}@{}}
\toprule
\textbf{Spec.} & \textbf{Accuracy} & \textbf{Precision} & \textbf{Recall} \\
\midrule
1 & 1 & 1 & 5.6 \\ 
2 & 0 & 0 & 0 \\ 
3 & 3.3 & 3.9 & 3 \\ 
4 & 1.1 & 1.8 & 11.1 \\ 
5 & 0.4 & 2.8 & -1 \\ 
6 & 0.6 & 0.4 & 2.4 \\ 
7 & 1.6 & 0.9 & 10.1 \\ 
8 & 0.8 & 0 & 2.6 \\ 
9 & -0.8 & -3.7 & -2.9 \\ 
10 & 0.4 & -2 & 5.8 \\ 
\midrule
           $\varnothing$ & 0.8 & 0.5 & 3.7  \\ 
\bottomrule
\end{tabular}
\end{table}

The results are shown in Table~\ref{tab:results:hyperparameters}. While the impact on accuracy and precision is minimal, recall can be improved significantly for some specifications. In particular, specifications 4 and 7 benefit here. While specification 4 was one of the negative outliers in Phase 1 (cf.~\ref{tab:results:crossproject-phase1}) and thus had much room for improvement, specification 7 had a better starting position ($62.0$), but improved significantly to $86.6$ in Phase 2. It is not exactly clear why these specifications benefited most from the tuned hyper-parameters and a detailed investigation is beyond the scope of this paper. However, if recall is an issue, improving the hyper-parameters can improve the results on average. Since this step is also relatively easy and quick to do with modern machine learning toolkits like scikit-learn, we recommend to investigate the potential improvements with the concrete dataset under consideration.

\subsubsection{Important Features}
\label{sec:results:features}

If a classifier is robust w.r.t.~different ways of specifying requirements, it will be easier to train a cross-project classifier. We thus set out to better
 understand which impact the heterogeneity of the requirement specifications has on the classifier and thus investigated the features the classifier uses to make a decision about the requirement. In our case, a feature is a word stem derived from the training data and deemed as relevant to the detection of a security requirement. We thus expect the most relevant features to contain stems such as ``access'' or ``encrypt''.
 
\begin{table}[tb]
\centering
\caption{The 25 top-ranked features used by the classifier when trained on different data sets along with their importance. Features in bold are not in the list for the full data set.}
\label{tab:results:features}
\begin{tabular}{@{}p{0.6cm}p{1.2cm}p{0.5cm}p{1.2cm}p{0.5cm}p{1.2cm}p{0.5cm}@{}}
\toprule
\textbf{Rank} & \textbf{Full data} & & \textbf{Spec.~3} && \textbf{Spec.~7}&\\
\midrule
1 & secur & 8.7 & author & 9.5 & secur & 10.8 \\ 
2 & user & 7.9 & user & 8.9 & access & 2.9 \\ 
3 & author & 7.8 & secur & 7.9 & password & 2.2 \\ 
4 & access & 4.1 & access & 2.7 & servic & 2.1 \\ 
5 & provid & 3.7 & provid & 2.4 & host & 2 \\ 
6 & host & 3.5 & abil & 1.8 & user & 1.9 \\ 
7 & servic & 2.4 & password & 1.5 & solut & 1.8 \\ 
8 & authent & 2.1 & \textbf{shall} & 1.2 & provid & 1.8 \\ 
9 & role & 1.9 & login & 1.1 & login & 1.5 \\ 
10 & protect & 1.7 & role & 0.9 & encrypt & 1.4 \\ 
11 & transact & 1.4 & log & 0.9 & \textbf{use} & 1.4 \\ 
12 & encrypt & 1.4 & authent & 0.9 & log & 1.4 \\ 
13 & solut & 1.3 & administr & 0.8 & authent & 1.2 \\ 
14 & password & 1.3 & encrypt & 0.8 & author & 1.2 \\ 
15 & abil & 1.1 & audit & 0.7 & polici & 1 \\ 
16 & polici & 1.1 & polici & 0.7 & administr & 1 \\ 
17 & audit & 1.1 & web & 0.6 & abil & 1 \\ 
18 & vulner & 0.8 & vulner & 0.6 & audit & 1 \\ 
19 & chang & 0.8 & \textbf{base} & 0.6 & vulner & 0.9 \\ 
20 & administr & 0.8 & \textbf{use} & 0.6 & protect & 0.9 \\ 
21 & log & 0.7 & \textbf{configur} & 0.6 & role & 0.8 \\ 
22 & activ & 0.7 & solut & 0.5 & web & 0.8 \\ 
23 & web & 0.7 & modifi & 0.5 & right & 0.7 \\ 
24 & login & 0.6 & \textbf{right} & 0.4 & \textbf{group} & 0.5 \\ 
25 & right & 0.6 & \textbf{integr} & 0.4 & \textbf{manag} & 0.5 \\ 
\bottomrule
\end{tabular}
\end{table}

The results in Table~\ref{tab:results:features} support this expectation. The most important features for the full data are based on generic security terms such as ``security'' and ``authorize''. These terms  are indeed strong indicators for a security-related requirement. Other, more specific terms such as ``password''or ``login'' score significantly lower and are less differentiating.

\paragraph{Impact of specification type} 
As mentioned in Sect.~\ref{sec:method}, we selected specifications independent of how they were formulated and with a significant spread in length. A typical requirement would, e.g., be formulated as a user story:
\\
\begin{aquote}{Specification 5}
As a user of a Query Broker enabler I would like to write my own libraries for accessing the content context data via a single unified API without taking care about the specifics of the internal data storage and DB implementations and interfaces.
\end{aquote}

This requirement is functional and not labelled as security-related. An example of a security-related requirement formulated as an instruction to potential vendors is as follows: 

\begin{aquote}{Specification 3}
The Vendor shall ensure that all data at rest and in motion is secure and encrypted in compliance with HIPAA and applicable State law.
\end{aquote}

Finally, our specifications include natural language requirements written in the ``the system shall'' style such as this functional requirement:

\begin{aquote}{Specification 7}
The system shall provide the ability to age and analyze accounts payable  e.g.\  open payables at end of month vs normal average payable balance
\end{aquote} 

However, we did not expect the structure of the requirements to make a significant difference to the classifier since it operates on term frequency. Our results in Table~\ref{tab:results:features} confirm this to a certain degree. The lists of features are ranked by the priority assigned by the classifier. The terms are similar in the different lists, but differ in priority.

To substantiate this assessment, we also ran experiments in which we only included specifications of the same type. Referring back to Table~\ref{tab:method:projects}, we, e.g., selected only specifications of type \emph{Request for Proposals} (RFP) and ran a k-fold cross-validation on them. Since the data for Backlog (BL) and System Requirement Specification (SRS) is too sparse, we only report on RFP here.

\begin{table}[tb]
\centering
\caption{Cross-project validation results of RFP requirements specifications' type.}
\label{tab:results:withinspec}
\begin{tabular}{@{}p{0.6cm}p{1cm}p{1cm}p{1cm}p{0.7cm}p{0.7cm}p{0.7cm}@{}}
\toprule
\multicolumn{4}{l}{} & \multicolumn{3}{l}{\textbf{f-measures}} \\
\textbf{Spec.} & \textbf{Accuracy} & \textbf{Precision} & \textbf{Recall} & \textbf{f${}_1$} & \textbf{f${}_{1/2}$} & \textbf{f${}_2$}\\
\midrule
3 & 78.9 & 88.8 & 71 & 78.9 & 84.6 & 74 \\ 
6 & 91.8 & 86.5 & 78 & 82.1 & 84.7 & 79.6 \\ 
7 & 87.1 & 75.2 & 42.5 & 54.3 & 65.2 & 46.5 \\ 
8 & 89.8 & 96.3 & 68.4 & 80 & 89 & 72.6 \\ 
10 & 93.9 & 78.9 & 57.7 & 66.7 & 73.5 & 61 \\ 
\midrule
           $\varnothing$  & 89.62 & 85.14 & 63.52 & 72.4 & 79.7 & 66.9 \\  
\bottomrule
\end{tabular}
\end{table}

Training only on the same type of requirement specification does not yield better classification results. This is demonstrated in Table~\ref{tab:results:withinspec} where the quality metrics for a 5-fold cross-validation of RFP specifications are shown. Compared with the results of the 10-fold cross-validation in Table~\ref{tab:results:crossproject-phase2}, results have consistently decreased. This might be related to the fact that the classifiers were trained on significantly less data. This interpretation is supported by the negative outlier specification 7: the classifier for this specification is trained on the least amount of requirements.

These results show that homogeneity in the training data is not a necessity for high-quality classification, at least with regard to the specification type. It also shows that the way the requirements are written does not play a significant role.

\paragraph{Stop words}
Table~\ref{tab:results:features} shows some aberrations when considering the specific specifications. For Specification 3, e.g., the term ``shall'' is ranked on position 8. This is due to the way the requirements are written (``the system shall'').
Another interesting term is ``user''. In particular requirements that are written as user stories might contain this term in all requirements, not only the security-related ones. Finally, the term ``vendor'' appears in some specifications, in particular ones of the RFP type and is not exclusive to security-related requirements there.

\begin{table}[tb]
\centering
\caption{Changes in quality metrics when extending the stop words list with ``user'', ``vendor'', and ``shall'' for each specification.}
\label{tab:results:stopwords}
\begin{tabular}{@{}p{0.6cm}p{1.2cm}p{1.2cm}p{1.2cm}@{}}
\toprule
\textbf{Spec.} & \textbf{Accuracy} & \textbf{Precision} & \textbf{Recall} \\
\midrule
1 & 0 & 0 & 0 \\ 
2 & -3.7 & -0.1 & -8.3 \\ 
3 & 0.6 & -2.3 & 3 \\ 
4 & 0 & 1.8 & 0 \\ 
5 & -0.5 & -5.8 & 3.9 \\ 
6 & -0.6 & -2.5 & 0 \\ 
7 & -2.1 & -4 & -8.9 \\ 
8 & 0 & 0 & -2.6 \\ 
9 & 0 & 0 & 0 \\ 
10 & -0.8 & 1.6 & -9.6 \\ 
\midrule
           $\varnothing$ & -0.7 & -1.1 & -2.3 \\ 
\bottomrule
\end{tabular}
\end{table}

Excluding such terms from the classification makes the results worse on average. Table~\ref{tab:results:stopwords} shows the results of an experiment where we included the terms ``user'', ``vendor'', and ``shall'' in the list of stop-words. This means that these terms are no longer accessible to the classifier to distinguish the two classes. While recall improved minimally for some specifications (3 and 5), we observed either no impact (because the words were not used in the specification) or a deterioration. This means that these words are indeed relevant for the classifier. At least when it comes to ``user'' this is also visible in the importance that the term carries in Table~\ref{tab:results:features}.

\paragraph{Security mechanisms}
The concrete list of features also depends on the security mechanisms in place. The term ``password'', e.g., is included in all three lists. However, a system does not necessarily have to be password protected. There would be differences in the list of terms if, e.g., a public-key infrastructure would be used. While there are no such issues in the specifications we used, we do observe that specific terminology, e.g., names of protocols are used without describing that they are security related. Specification 10, e.g., contains the following two requirements: \inlinequote{The solution should support LDAPv3} and  \inlinequote{The solution should support Kerberos}. The fact that LDAP and Kerberos are used for authentication and are therefore security-related is not visible from these descriptions. Such requirements can negatively affect recall since they might not be picked up by a classifier who was not trained on requirements that explicitly contained the connection to security and thus has the respective terms in the list of features.

A similar problem is the use of security standards by name or reference without a description. An example is the requirement \inlinequote{Hosting Service Provider will possess an ISO 27002 Certificate of Conformance or equivalent certifications} from specification 3 and the explicit mention of HIPAA (a privacy standard for patient data) in specifications 3 and 6. The use of security-related terms not explicitly tied to security is also present in specification 3 in the requirement \inlinequote{Solutions will have a fraud detection function}.



\paragraph{Other misclassifications}

When analysing the results of the classification, we see several other misclassifications that have a negative impact on precision and recall. The naming of applications or modules in the system influences precision, e.g., when a part of the system is called ``the secure analytical environment'' as in specification 3. Every time a requirement refers to that module, there is a chance that the classifier will tag this as a security requirement because of the term ``secure''. Likewise, when requirements refer to security other than cyber-security (e.g., in specifictation 10: \inlinequote{The system should be able to manage and contain data on different employee categories: Internal: Statutory  SNE  Interim  Trainee ; External: Consultants  Security Personnel}), the requirement might be misclassified as a security-related requirement.

Other misclassifications can be observed in cases where requirements describe a certain aspect such as a login page and how a user interacts with it or include a sequence of events that includes a security-related action (e.g., ``After the user login he/she may submit a request'' in specification 2). Both of these issues affect precision since such requirements might be falsely classified as security-related.

\section{Feasibility of Cross-Project Classification}
\label{sec:feasibility}
Our results indicate that a classifier that has been trained on other projects can be useful for an initial classification of a large set of requirements. However, some initial manual effort is required to achieve sufficiently high precision and recall. Perfect scores are unobtainable due to the inherent ambiguity in the requirements. We discuss these aspects in the following in order to provide guidelines for practitioners who are thinking about labelling their requirements, e.g., in preparation for the creation of security assurance cases.

\subsection{Selection of training data}

Suitable training data should include use of relevant security mechanisms and standards. If the classifier is trained with training data that uses password authentication but the requirements to be labelled do not, the risk of false negatives increases and recall declines. Since we have randomly selected the first ten requirement specifications we found and that adhered to our criteria, we have presented the classifier with a worst-case scenario.

However, in many cases the different requirement specifications for a product can be heterogeneous themselves. A vehicle, e.g., consists of many different components, developed by different teams, vendors, and engineers from different disciplines, using different development paradigms. Some parts might be described as a specification handed over to a vendor (as our specification 3), others as item definitions according to ISO~26262~\cite{iso26262}, and yet others as user stories that resided in a product backlog. If this is the case, a heterogeneous training set is an advantage since the resulting classifier can be used to label all requirements. An alternative would be to train separate requirements for the different types. However, as we have shown in Section~\ref{sec:results:phase3}, a classifier trained only on Request for Proposal (RFP) requirement specification fared worse than the classifier trained on all types of specifications.

These results are a bit surprising when comparing them with related work. Knauss et al.~\cite{knauss2011}, e.g., found that when they trained a Bayesian classifier on two different projects, it fared poorly when applied to a third project. We hypothesise that a certain minimal number of specifications is necessary for the classifier to become general enough. If the classifier has been trained on a sufficient number of variations on a certain term, the likelihood that it will pick up a certain phrasing increases, thus improving recall. An example from our data is the use of the term ``sensitive data'' in specification 10 which is not used in any other specification. We thus recommend to select training data that is as heterogeneous as possible.



\subsection{Manual Labelling}

High-quality labelling of the training data has the biggest impact on classification quality. As discussed in Section~\ref{sec:results:phase2}, results improved significantly when the training data included consistent labels for security requirements. This extra effort needs to be weighed against the benefits. Especially in cases where the unlabelled data set to which the classifier should be applied is very large, investing time in labelling a comparatively small set of training data might be significantly faster than a full manual labelling.

In order to revise the labelling for Phase 2 of the validation, one of the authors of this study spent approximately 15 hours on the 3003 requirements included in the ten specifications. Overall, we estimate that the effort of preparing the classifier, collecting publicly available data, cleaning up the data, relabelling it, and running the classification took approximately one entire work week for one person. When compared to the effort that can be spent on manually labelling reasonably large specifications, we believehttps://www.overleaf.com/project/5e3c0ecc7896760001769b31 this to be a worthwhile investment.


\subsection{Interpreting the results}

Our original goal was to determine which performance a classifier for security-related requirements can achieve when trained on data from other projects (\textbf{RQ1}). Our results shown in Table~\ref{tab:results:crossproject-phase2} indicate that a precision above $85\%$ can be achieved for most specifications while recall varies significantly between $65\%$ and $90\%$. These numbers were achieved after manually revising the labelling of the training data.

We have also shown that the tuning of hyper-parameters has an impact mostly on recall, that manipulating the stop-words list deteriorates results, and that a more heterogeneous set of training data is advantageous. This answers \textbf{RQ2} about the impact of different factors on classification performance.

We believe that the performance we were able to achieve will be sufficient to support engineers in the initial labelling of a large set of requirements, in particular if the training data is heterogeneous enough and contains a large enough number of security-related requirements. All of our findings indicate that the more heterogeneity in the training data and the more requirements in the training data, the better the performance will be. Our findings also indicate, however, that some manual effort is necessary to ensure the quality of the labelling in the training data is sufficient. How much time an organisation will be willing to invest might depend on the number of requirements that have to be labelled --- spending 15 hours on labelling 3000 requirements might be significantly less effort than manually labelling tens of thousands of requirements.

We have also shown the limitations of automated classification. In particular, short requirements that name specific security mechanisms, standards or unconventional technology will be difficult to classify automatically. Likewise, architectural elements with names that contain security related terms or requirements that mention security-related terms without referring to cyber-security have a negative impact on precision and recall. We have listed a number of such examples in Section~\ref{sec:results:phase3}. They are partially responsible for the numbers that we see and are, in practice, unavoidable. 

A possible extension of our approach is to use online learning to incorporate manually curated data from the project under investigation. This way, the initial learning on freely available data can be complemented with parts of the project's actual requirement specification that have already been labelled. There are a number of random forest variants for online learning (see, e.g., \cite{saffari2009line, lakshminarayanan2014mondrian}) that can be used for this purpose. Alternatively, a Bayesian classifier such as the one used by Knauss et al.~\cite{knauss2011} can be used. Such an approach would continuously improve predictive performance while the engineers revise the labelling suggested by an initial classification.

\section{Conclusion}
\label{sec:conclusion}
We have shown that it is feasible to train a classifier for security-related requirements on publicly available data and achieve a satisfactory classification performance when applying it to a new specification after manually revising the labelling of the training data. Our results furthermore indicate that heterogeneity in the training data is an advantage in cross-project classification and that interventions such as changing the list of stop words have no positive effect on performance.

In our future work, we will use these results in the context of security assurance cases. We will support our industrial partners that need to construct such cases by providing our classifier to them in order to create an initial labelling of their requirement specifications as a foundation for defining the claims. We will also investigate the possibility to extend our approach towards online learning to seamlessly incorporate parts of the requirement specifications that have already been labelled and vetted.

\clearpage 
\bibliographystyle{ieeetran}
\bibliography{bibliography}

\begin{thebibliography}{10}
\providecommand{\url}[1]{#1}
\csname url@samestyle\endcsname
\providecommand{\newblock}{\relax}
\providecommand{\bibinfo}[2]{#2}
\providecommand{\BIBentrySTDinterwordspacing}{\spaceskip=0pt\relax}
\providecommand{\BIBentryALTinterwordstretchfactor}{4}
\providecommand{\BIBentryALTinterwordspacing}{\spaceskip=\fontdimen2\font plus
\BIBentryALTinterwordstretchfactor\fontdimen3\font minus
  \fontdimen4\font\relax}
\providecommand{\BIBforeignlanguage}[2]{{%
\expandafter\ifx\csname l@#1\endcsname\relax
\typeout{** WARNING: IEEEtran.bst: No hyphenation pattern has been}%
\typeout{** loaded for the language `#1'. Using the pattern for}%
\typeout{** the default language instead.}%
\else
\language=\csname l@#1\endcsname
\fi
#2}}
\providecommand{\BIBdecl}{\relax}
\BIBdecl

\bibitem{sac}
J.~{Knight}, ``The importance of security cases: Proof is good, but not
  enough,'' \emph{IEEE Security Privacy}, vol.~13, no.~4, pp. 73--75, July
  2015.

\bibitem{jeepHack}
H.~{Kwon}, S.~{Lee}, J.~{Choi}, and B.~{Chung}, ``Mitigation mechanism against
  in-vehicle network intrusion by reconfiguring ecu and disabling attack
  packet,'' in \emph{2018 International Conference on Information Technology
  (InCIT)}, Oct 2018, pp. 1--5.

\bibitem{iso21434}
{Technical Committee ISO/TC 22/SC 32 Electrical and electronic components and
  general system aspects}, ``{ISO/SAE DIS 21434 [SAE], Road vehicles ---
  Cybersecurity engineering},'' p. 101, 2020, preview.

\bibitem{cnn}
J.~{Winkler} and A.~{Vogelsang}, ``Automatic classification of requirements
  based on convolutional neural networks,'' in \emph{2016 IEEE 24th
  International Requirements Engineering Conference Workshops (REW)}, Sep.
  2016, pp. 39--45.

\bibitem{firesmith2003}
D.~Firesmith \emph{et~al.}, ``Engineering security requirements.''
  \emph{Journal of object technology}, vol.~2, no.~1, pp. 53--68, 2003.

\bibitem{knauss2011}
E.~Knauss, S.~Houmb, K.~Schneider, S.~Islam, and J.~J{\"u}rjens, ``Supporting
  requirements engineers in recognising security issues,'' in
  \emph{International Working Conference on Requirements Engineering:
  Foundation for Software Quality}.\hskip 1em plus 0.5em minus 0.4em\relax
  Springer, 2011, pp. 4--18.

\bibitem{kurtanovic2017}
Z.~Kurtanovi{\'c} and W.~Maalej, ``Automatically classifying functional and
  non-functional requirements using supervised machine learning,'' in
  \emph{2017 IEEE 25th International Requirements Engineering Conference
  (RE)}.\hskip 1em plus 0.5em minus 0.4em\relax IEEE, 2017, pp. 490--495.

\bibitem{schatz2017}
D.~Schatz, R.~Bashroush, and J.~Wall, ``Towards a more representative
  definition of cyber security,'' \emph{Journal of Digital Forensics, Security
  and Law}, vol.~12, no.~2, pp. 53--74, 2017.

\bibitem{samonas2014}
S.~Samonas and D.~Coss, ``The cia strikes back: Redefining confidentiality,
  integrity and availability in security.'' \emph{Journal of Information System
  Security}, vol.~10, no.~3, 2014.

\bibitem{iso27001}
{International Organization for Standardization}, ``{ISO/IEC 27001 INFORMATION
  SECURITY MANAGEMENT, 1st Edition},'' {Geneva, Switzerland}, {2013}.

\bibitem{GSN_standard}
G.~C. S.~W. Group, ``Gsn community standard,'' \emph{Available at
  www.goalstructuringnotation.info/}, 2011.

\bibitem{alexander2011}
R.~Alexander, R.~Hawkins, and T.~Kelly, ``Security assurance cases: motivation
  and the state of the art,'' \emph{High Integrity Systems Engineering
  Department of Computer Science University of York Deramore Lane York YO10
  5GH}, 2011.

\bibitem{goodenough2007}
J.~Goodenough, H.~Lipson, and C.~Weinstock, ``Arguing security-creating
  security assurance cases,'' \emph{rapport en ligne (initiative build
  security-in du US CERT), Universit{\'e} Carnegie Mellon}, 2007.

\bibitem{38_agudo2009}
I.~Agudo, J.~L. Vivas, and J.~L{\'o}pez, ``Security assurance during the
  software development cycle,'' in \emph{Proceedings of the International
  Conference on Computer Systems and Technologies and Workshop for PhD Students
  in Computing}.\hskip 1em plus 0.5em minus 0.4em\relax ACM, 2009, p.~20.

\bibitem{54_haley2005}
C.~B. Haley, J.~D. Moffett, R.~Laney, and B.~Nuseibeh, ``Arguing security:
  Validating security requirements using structured argumentation,'' 2005.

\bibitem{55_calinescu2017}
R.~Calinescu, D.~Weyns, S.~Gerasimou, M.~U. Iftikhar, I.~Habli, and T.~Kelly,
  ``Engineering trustworthy self-adaptive software with dynamic assurance
  cases,'' \emph{IEEE Transactions on Software Engineering}, vol.~44, no.~11,
  pp. 1039--1069, 2017.

\bibitem{friedman2001}
J.~Friedman, T.~Hastie, and R.~Tibshirani, \emph{The elements of statistical
  learning}.\hskip 1em plus 0.5em minus 0.4em\relax Springer series in
  statistics New York, 2001, vol.~1, no.~10.

\bibitem{docRet}
K.~{Chindamaikul}, T.~{Takai}, and H.~{Iida}, ``Retrieving information from a
  document repository for constructing assurance cases,'' in \emph{2014 IEEE
  International Symposium on Software Reliability Engineering Workshops}, Nov
  2014, pp. 198--203.

\bibitem{hawkins2015}
R.~Hawkins, I.~Habli, D.~Kolovos, R.~Paige, and T.~Kelly, ``Weaving an
  assurance case from design: a model-based approach,'' in \emph{2015 IEEE 16th
  International Symposium on High Assurance Systems Engineering}.\hskip 1em
  plus 0.5em minus 0.4em\relax IEEE, 2015, pp. 110--117.

\bibitem{shrott2015}
C.~Shortt and J.~Weber, ``Hermes: A targeted fuzz testing framework,'' 09 2015.

\bibitem{tippenhauer2014}
N.~O. Tippenhauer, W.~G. Temple, A.~H. Vu, B.~Chen, D.~M. Nicol, Z.~Kalbarczyk,
  and W.~H. Sanders, ``Automatic generation of security argument graphs,'' in
  \emph{2014 IEEE 20th Pacific Rim International Symposium on Dependable
  Computing}.\hskip 1em plus 0.5em minus 0.4em\relax IEEE, 2014, pp. 33--42.

\bibitem{xu2007}
Q.~Xu, R.~Jiao, X.~Yang, M.~Helander, H.~Khalid, and O.~Anders, ``Customer
  requirement analysis based on an analytical kano model,'' in \emph{2007 IEEE
  International Conference on Industrial Engineering and Engineering
  Management}.\hskip 1em plus 0.5em minus 0.4em\relax IEEE, 2007, pp.
  1287--1291.

\bibitem{ott2014}
D.~Ott and F.~Houdek, ``Automatic requirement classification: Tackling
  inconsistencies between requirements and regulations,'' \emph{International
  Journal of Semantic Computing}, vol.~8, no.~01, pp. 47--65, 2014.

\bibitem{cleland2007}
J.~Cleland-Huang, R.~Settimi, X.~Zou, and P.~Solc, ``Automated classification
  of non-functional requirements,'' \emph{Requirements Engineering}, vol.~12,
  no.~2, pp. 103--120, 2007.

\bibitem{scikit-learn}
F.~Pedregosa, G.~Varoquaux, A.~Gramfort, V.~Michel, B.~Thirion, O.~Grisel,
  M.~Blondel, P.~Prettenhofer, R.~Weiss, V.~Dubourg, J.~Vanderplas, A.~Passos,
  D.~Cournapeau, M.~Brucher, M.~Perrot, and E.~Duchesnay, ``Scikit-learn:
  Machine learning in {P}ython,'' \emph{Journal of Machine Learning Research},
  vol.~12, pp. 2825--2830, 2011.

\bibitem{manning2008introduction}
C.~D. Manning, P.~Raghavan, and H.~Sch{\"u}tze, \emph{Introduction to
  information retrieval}.\hskip 1em plus 0.5em minus 0.4em\relax Cambridge
  University Press, 2008.

\bibitem{chawla2002smote}
N.~V. Chawla, K.~W. Bowyer, L.~O. Hall, and W.~P. Kegelmeyer, ``Smote:
  synthetic minority over-sampling technique,'' \emph{Journal of artificial
  intelligence research}, vol.~16, pp. 321--357, 2002.

\bibitem{ho1995random}
T.~K. Ho, ``Random decision forests,'' in \emph{Proceedings of 3rd
  international conference on document analysis and recognition}, vol.~1.\hskip
  1em plus 0.5em minus 0.4em\relax IEEE, 1995, pp. 278--282.

\bibitem{platt1998svm}
J.~Platt, ``Sequential minimal optimization: A fast algorithm for training
  support vector machines,'' 1998.

\bibitem{maron1961nb}
M.~E. Maron, ``Automatic indexing: an experimental inquiry,'' \emph{Journal of
  the ACM (JACM)}, vol.~8, no.~3, pp. 404--417, 1961.

\bibitem{kappa}
\BIBentryALTinterwordspacing
J.~Cohen, ``A coefficient of agreement for nominal scales,'' \emph{Educational
  and Psychological Measurement}, vol.~20, no.~1, pp. 37--46, 1960. [Online].
  Available: \url{https://doi.org/10.1177/001316446002000104}
\BIBentrySTDinterwordspacing

\bibitem{kappa1}
T.~O'Dowd, R.~West, P.~Winterburn, and M.~Hewlins, ``Evaluation of a rapid
  diagnostic test for bacterial vaginosis,'' \emph{BJOG: An International
  Journal of Obstetrics \& Gynaecology}, vol. 103, no.~4, pp. 366--370, 1996.

\bibitem{kappa2}
B.~Szebenyi, A.~P. Hollander, P.~Dieppe, B.~Quilty, J.~Duddy, S.~Clarke, and
  J.~R. Kirwan, ``Associations between pain, function, and radiographic
  features in osteoarthritis of the knee,'' \emph{Arthritis \& Rheumatism:
  Official Journal of the American College of Rheumatology}, vol.~54, no.~1,
  pp. 230--235, 2006.

\bibitem{hawkins2004}
D.~M. Hawkins, ``The problem of overfitting,'' \emph{Journal of chemical
  information and computer sciences}, vol.~44, no.~1, pp. 1--12, 2004.

\bibitem{imbalanced}
S.~Kotsiantis, D.~Kanellopoulos, and P.~Pintelas, ``Handling imbalanced
  datasets: A review,'' \emph{GESTS International Transactions on Computer
  Science and Engineering}, vol.~30, pp. 25--36, 11 2005.

\bibitem{iso26262}
{Technical Committee ISO/TC 22/SC 32 Electrical and electronic components and
  general system aspects}, ``{ISO 26262-1:2018, Road vehicles --- Functional
  safety},'' 2018.

\bibitem{saffari2009line}
A.~Saffari, C.~Leistner, J.~Santner, M.~Godec, and H.~Bischof, ``On-line random
  forests,'' in \emph{IEEE 12th International Conference on Computer Vision
  (ICCV) workshops}.\hskip 1em plus 0.5em minus 0.4em\relax IEEE, 2009, pp.
  1393--1400.

\bibitem{lakshminarayanan2014mondrian}
B.~Lakshminarayanan, D.~M. Roy, and Y.~W. Teh, ``Mondrian forests: Efficient
  online random forests,'' in \emph{Advances in Neural Information Processing
  Systems}, 2014, pp. 3140--3148.

\end{thebibliography}

\end{document}